\documentclass{article}
\usepackage[utf8]{inputenc}
\usepackage{arxiv}
\usepackage{cite}
\usepackage{hyperref}

\usepackage{amsmath,amssymb,amsfonts}
\usepackage{ifthen}
\usepackage{algorithm2e}

\usepackage{graphicx}
\usepackage{algorithmic}
\usepackage{float}
\usepackage{textcomp}
\usepackage{centernot}
\usepackage[braket, qm]{qcircuit}
\usepackage{braket}
\usepackage{physics}
\def\BibTeX{{\rm B\kern-.05em{\sc i\kern-.025em b}\kern-.08em
    T\kern-.1667em\lower.7ex\hbox{E}\kern-.125emX}}
\usepackage{multicol}
\usepackage{lipsum}
\usepackage{subcaption}
\usepackage{wrapfig}
\usepackage{tabularx}
\usepackage{cleveref}

\hypersetup{
    colorlinks,
    linkcolor={black},
    citecolor={black},
    urlcolor={blue}
}
\usepackage{tikz}

\usetikzlibrary{intersections}

\usepackage{amsmath}
\DeclareUnicodeCharacter{2212}{−}

\usetikzlibrary{patterns,shapes.arrows}

\usetikzlibrary{shapes.geometric}
\usetikzlibrary{arrows.meta,
                calc,
                positioning,
                quotes}
\definecolor{infineon_dark_green}{HTML}{0A8276}
\definecolor{infineon_purple}{HTML}{9C216E}
\definecolor{infineon_light_green}{HTML}{9BBA43}
\definecolor{infineon_orange}{HTML}{F97414}
\usepackage{mathtools}
\usepackage{listings}
\usepackage[numbers]{natbib}
\usepackage{url}
\usepackage{booktabs} 

\usepackage{array}
\newcommand{\PreserveBackslash}[1]{\let\temp=\\#1\let\\=\temp}
\newcolumntype{C}[1]{>{\PreserveBackslash\centering}p{#1}}
\tikzset{SubCaption/.style={
text width=2in,yshift=-3mm, align=center,anchor=north, 
}}

\usepackage[none]{hyphenat}
\usepackage{flushend}

\title{Grover adaptive search with problem-specific state preparation}
\date{} 
\author{
Maximilian Hess\thanks{Infineon Technologies AG, Neubiberg, Germany, maximilian.hess@infineon.com}, \and 
Lilly Palackal\footnotemark[1], \and 
Abhishek Awasthi\thanks{BASF Digital Solutions GmbH, Ludwigshafen, Germany}, \and 
Peter J. Eder\thanks{Siemens AG, Munich, Germany}, \and
Manuel Schnaus\footnotemark[1], \and 
Laurin Demmler\footnotemark[1], \and 
Karen Wintersperger\footnotemark[3],\and 
Joseph Doetsch\thanks{
Lufthansa Industry Solutions GmbH, Norderstedt, Germany}\\
\textit{This paper was developed within the Quantum Technology and Application Consortium (QUTAC).}\thanks{QUTAC: info@qutac.de}
}

\begin{document}
\maketitle

\begin{abstract}
Grover's search algorithm is one of the basic building block in the world of quantum algorithms. Successfully applying it to combinatorial optimization problems is a subtle challenge. As a quadratic speedup is not enough to naively search an exponentially large space, the search has to be complemented with a state preparation routine which increases the amplitudes of promising states by exploiting the problem structure. In this paper, we build upon previous work~\cite{state_preparation} to construct heuristic state preparation routines for the Traveling Salesperson Problem (TSP), mimicking the well-known classical Lin-Kernighan heuristic. With our heuristic, we aim to achieve a reasonable approximation ratio with only a polynomial number of Grover iterations.
Further, we compare several algorithmic settings relating to termination criteria and the choice of Grover iterations when the number of marked solutions is unknown.
\end{abstract}

\section{Introduction}

Grover's algorithm~\cite{grover_algorithm} is one of the oldest and most general achievements in the field of quantum algorithms. The algorithm searches an unsorted space for elements identified by a so-called oracle offering a quadratic speed-up (in terms of oracle calls) compared to brute-force search. When trying to apply this search algorithm to the domain of combinatorial optimization multiple challenges arise. 
The first is to define the criterion to be checked by the oracle. Given a candidate solution, it is in general not possible to efficiently tell whether the candidate solution optimizes the problem.
To circumvent this issue, previous work proposed to instead focus on iteratively searching for improvements with respect to a reference solution instead of searching for the best solution from the beginning~\cite{grover_adaptive_search, fixed_point_quantum_search,durr1999quantumalgorithmfindingminimum}. The initial reference is typically obtained using an efficient heuristic (e.g., a greedy algorithm), after which the Grover algorithm searches for all states better than the reference. For this search, the oracle can be easily defined~\cite{grover_adaptive_search} by marking all states whose cost function is lower than the reference value. After sampling a state improving the cost function, this state is chosen as the new reference and the process is repeated. When the Grover search cannot find an improvement, the algorithm returns the current solution.
In this work, we address two key issues around this concept: The first concerns the details of the Grover search itself. The optimization application differs from the textbook case as we usually do not know how many ``good" states there are. Typically, this problem is addressed by a randomization of the number of Grover iterations~\cite{Boyer_1998_tightBoundsQuantumSearch,durr1999quantumalgorithmfindingminimum} to be performed. We showcase a few subtleties in how exactly this randomization takes place and investigate if the same result could be achieved without randomness.
The second issue addressed in this work is the state preparation preceding the Grover/Amplitude Amplification procedure. If one starts the algorithm for the solution of an optimization problem with $n$ binary variables in the uniform superposition over all $2^n$ bit strings, there is no escaping an exponential run time if the goal is to measure one of exponentially few solutions. For this reason, we assert that any successful application of quantum search algorithms in combinatorial optimization must come with a problem-specific state preparation routine with the purpose of filtering out infeasible states and/or skewing the superposition towards certain reference states. This concept has already been demonstrated for the knapsack problem in~\cite{wilkening_2023_QTGKnapsack}.
We aim to demonstrate a similar method for the well known Traveling Salesperson Problem (TSP) whose constraint structure is fundamentally different from the linear inequalities defining the knapsack problem. To reach our objective, we combine a state preparation routine for states representing Hamiltonian cycles from~\cite{state_preparation} with inspiration taken from the Lin-Kernighan heuristic. More precisely, at every improvement step, we prepare a superposition of routes that differ only in a limited number of positions from the incumbent solution. We employ a benchmarking scheme along the lines of~\cite{Cade_2023} and obtain relevant measures of quality for different algorithm settings. The hybrid benchmarking scheme allows us to analyze instances exceeding $100$ binary variables which would not be possible using other quantum algorithms such as the Quantum Approximate Optimization Algorithm (QAOA)~\cite{farhi2014quantumapproximateoptimizationalgorithm}.

\section{Methods and Related Work}
Attempts at applying quantum search algorithms in the optimization of functions date back to the 1990s. In~\cite{durr1999quantumalgorithmfindingminimum} the mechanism of iteratively searching for improvements with respect to an incumbent solution is outlined. The quantum search subroutine for an unknown number of good states in turn comes from~\cite{Boyer_1998_tightBoundsQuantumSearch} where the randomization of the choice of Grover iterations is proposed and proven to preserve a quadratic speedup over classical search.
Further options in searching a space with an unknown fraction of positive outcomes are discussed by Yoder, Low and Chuang~\cite{fixed_point_quantum_search}. They remove the randomness by an algorithm which guarantees to measure a good solution with a specifiable probability $1 - \delta$.
However, this guarantee comes at the price of an additional scaling which punishes small choices of $\delta$.
In~\cite{Gilliam_2021_GroverAdaptiveSearch}, the authors put together these well-known algorithmic building block into a framework dubbed ``Grover Adaptive Search" (GAS) and specifically deal with the efficient construction of oracle operators for QUBO problems.
On the side of combinatorial optimization, concrete applications of Grover Adaptive search have been proposed in~\cite{wilkening_2023_QTGKnapsack} and~\cite{Lu_2025_GASforVehicleRouting} for knapsack and vehicle routing problems respectively. As it is central to our work, we reproduce the GAS algorithm~\cite{grover_adaptive_search} below.

\begin{algorithm}[H]
\caption{Grover Adaptive Search in the orignial version. Whether or not the algorithm finds the optimal solution depends largely on when the search is terminated.}\label{alg:Grover Adaptive Search}
\KwData{$f: X \rightarrow \mathbb{R}$, $\lambda>1$}
\KwResult{Solution candidate $x$ and corresponding value $y = f(x)$}
{Compute $x_0$ from a heuristic and set $x=x_0$, $y=f(x_0)$\;
Set $k=1$\;}
\While{Termination condition is not met}    
{Randomly select $n_{\text{grover}}$ from $\{0, \dots ,\lceil k-1\rceil\}$;
Apply Grover with $n_{\text{grover}}$ iterations of oracle $O_{y}$ and diffusion operator $D$\; 
Measure solution candidate $z$\;
\If{$f(z) < y$}
    {Set new threshold $y=f(z)$ and incumbent solution $x=z$\;
    Reset $k=1$\;}
\Else
    {Set $k= \lambda\cdot k$\;}
}
\Return $x,y$
\end{algorithm}

The state preparation methods used in this paper rely on the methods developed by Baertschi and Eidenbenz~\cite{state_preparation} where a quantum algorithm which prepares superpositions of feasible states for the TSP is presented. Its original purpose was the construction of constraint-preserving mixers for variational algorithms.
This state preparation routine, in turn, relies crucially on the efficient preparation of Dicke states, based on work by the same authors~\cite{dicke_states}.
The idea to introduce a biasing  factor into the state preparation rather than aiming for an equal superposition is inspired by work on the knapsack problem~\cite{wilkening_2023_QTGKnapsack}, where a very performant quantum algorithm was presented on the basis of skewing the initial state towards a certain reference state.
Further inspiration comes from the classical optimization of the TSP where the $k$-opt and Lin-Kernighan heuristics~\cite{lin_kernighan} have long been a key component of state-of-the-art solvers.

\section{Grover Adaptive Search for the TSP}

\subsection{TSP: problem definition}
Originally formulated by William Rowan Hamilton and Thomas Kirkman~\cite{historyTSP} as a recreational puzzle, the Traveling Salesperson Problem is an essential problem in combinatorial optimization. It formalizes the problem of a traveler, who means to visit $n$ cities and returning to the city they originally left from, while never visiting a city twice and minimizing the total distance traveled. The problem is evidently influenced by pre-modern life and was first mentioned, albeit not in a mathematically descriptive manner in 1832\cite{history2TSP}.

Formally, the problem has the following setting: Given a graph $G$, defined by its set of vertices $V$ of size $n=|V|$ and edges $E$ (since the graph is complete, $|E|=\frac{n(n-1)}{2}$), such that $G = (V, E)$, we define each edge $e_{m,n} \in E$ to carry a weight $d_{m,n}$. In the original interpretation of the problem, $d_{m,n}$ corresponds to the distance between the vertices (cities) $v_m$ and $v_n$. A possible path through the graph can then be encoded fully in $N=n^2$ binary decision variables. Each binary variable $x_{i,t} \in \{0,1\}$ decides whether vertex $v_i \in V$ is visited at timestep $t \leq n$. Hence each vertex $v_i$ is assigned exactly $n$ variables, one for each timestep of the tour.

A feasible route in the graph must follow some obvious rules. 
\begin{enumerate}
    \item Each city must be visited exactly once in a feasible tour.
    \item At each point in time $t \leq n$, exactly one city is visited.
    \item There must be exactly $n$ total time steps (i.e. no timestep may be skipped). \cite{lin_kernighan}
\end{enumerate}

Mathematically these constraints define a set of feasible bit strings $F$ that describe allowed tours in the graph:
\begin{align}
    F = \Bigl\{x \in \{0,1\}^{N} : &\sum_{t=1}^{n} x_{i,t} = 1, \forall   i = 1,...,n \notag\\
    \text{and}\hspace{1cm} &\sum_{i=1}^{n} x_{i,t} = 1, \forall t = 1,...,n \Bigr\}.
\end{align}
To simplify notation, we define the cost (or distance) of some bit string $x\in F$ as $d(x)$ as:
\begin{equation}
    d(x) = \sum_{t=1}^{n} \sum_{i=1}^{n} \sum_{j=1}^{i-1} d_{i,j} x_{i,t} x_{j,(t \bmod n)+1},
\end{equation}
in which the modulo ensures that the route is a closed loop through the graph.
The TSP problem is solved by a bit string (or several if the optimal space is degenerate) that satisfies:
\begin{equation}
    x^* \in \arg\min_{x \in F} d(x).
\end{equation}

\subsection{Initial state preparation}\label{subsec:state_prep}

In this work, we propose a heuristic state preparation for the TSP, significantly reducing the search space while still producing high-quality solutions. Our state preparation is inspired by the classical Lin-Kernighan TSP heuristic~\cite{lin_kernighan}.
Our expectation is that by a combination of a Lin-Kernighan like state preparation and an amplitude amplification protocol, we can achieve a quadratic speedup over classical heuristics.
In the Lin-Kernighan heuristic, a chain of consecutive edges of variable length is exchanged by a set of edges not contained in the current route. 
Since the one-hot encoding of the TSP uses a node-based representation of a route, we exchange consecutive nodes in the current reference solution $x$ with other nodes in the graph. For each length of the exchange chain, we define a state preparation that creates a superposition of all solutions where those nodes are replaced. 
Additionally, we try starting the exchange chain at each step of the route to enable all possible node exchanges. Given a starting index $i_0$ and an exchange chain length $l$, we are interested in the set of solutions obtained by swapping $l$ indices of the original solutions such that no two indices are involved in more than one swap.
When sampling a new state, it can be assumed without loss of generality that the exchange chain starts at the first step in the route. 
This is possible due to the cyclic property of a route. To represent an exchange chain not starting at the beginning of the route, we can shift the start of the route so the route's starting point corresponds to the start of the chain.\\

\begin{algorithm}
\caption{Quantum search algorithm with a state preparation inspired by the Lin-Kernighan algorithm. The function $\text{sample}$ randomly samples a state switching a chain of a certain length of the reference starting at a specific TSP step after applying $n_{\text{grover}}$ Grover iterations. $d(x)$ describes the TSP distance of a solution $x$.}\label{alg:qts}
\KwData{Initial reference $x$, chain length limit $l_\text{max}$, Number of TSP cities $N$}
\KwResult{TSP Route $x$, estimated Grover iterations $k_\text{total}$}
$l_\text{chain} \gets 0$\;
$\text{improvement} \gets False$\;
$k_\text{total} \gets 0$\;
$\lambda \gets 5/4$\;
\While{$\text{improvement}$ or $l_\text{chain} < l_\text{max}$}{
$\text{improvement} \gets False$\;
$l_\text{chain} \gets l_\text{chain} + 1$\;
$i \gets 0$\;
$K \gets \min(N^{2}, N^{l_\text{chain}})$\;
\While{$i < N - 1$}{
$i \gets i + 1$\;
$k \gets 0$\;
$l \gets 0$\;
\While{$k < K$ and $k_{\text{total}} \leq 5N^3$}{
sample $n_{\text{grover}}$ uniformly at random from $\{0,\dots,\lambda^l\}$\;
$x_\text{new} \gets \text{sample}(x, l_\text{chain}, i, n_{\text{grover}})$\;
    \If{$d(x_\text{new}) < d(x)$}{
    $\text{improvement} \gets True$\;
    $k \gets k + n_{\text{grover}}$\;
    $l \gets l+1$ \;
    $x \gets x_\text{new}$\;
    $l_{\text{chain}} \gets 1$\;
    }
    $k_\text{total} \gets k_\text{total} + n_{\text{grover}}$\;
}
}
}
\end{algorithm}
The total algorithm for the tree search with creating a TSP with an estimated number of Grover iterations is described in Algorithm~\ref{alg:qts}. In this algorithm, we choose to expand the size of an exchange chain if the previous chains find an improvement. Since small exchange chain may not find any improvements depending on the heuristic used for the initial reference state, we introduce a parameter $l_\text{min}$. This parameter represents the size of exchanges up to which the algorithm has to check for improvements.\\
For each exchange chain at each location in the route, we set a maximum number of random samples until which an improvement has to be found. We adapt this limit depending on the current size of the exchange chain. With a TSP of size $n$ and a chain length of size $l$, the total number of samples is close to $\frac{(n - 1)!}{(n - 1 - l)!}$. Assuming one of those samples creates an improvement and the samples are drawn uniformly, limiting the number of iterations to $n^{l + 1}$ leads to a probability close to zero of not sampling that state for increasing values of $n$ and $l$.

In the small sized instances which we analyzed for this work, we observed that the exchange length to successfully move away from the greedy starting solution would often be so large that essentially all vertices are re-assigned at every improvement step. 
However, for larger instances beyond the scope of our simulation protocol, we expect to see an effect for $l << n$.

In order to create the samples, it is necessary to create a quantum state which switches out the nodes in the exchange chain while maintaining all other nodes in the TSP route. B{\"a}rtschi et al. have previously created a state preparation which encodes all feasible TSP solutions~\cite{state_preparation}. Here, we use that circuit structure to enable switching nodes in the exchange chain while setting all other nodes to values depending on the previous swaps. 
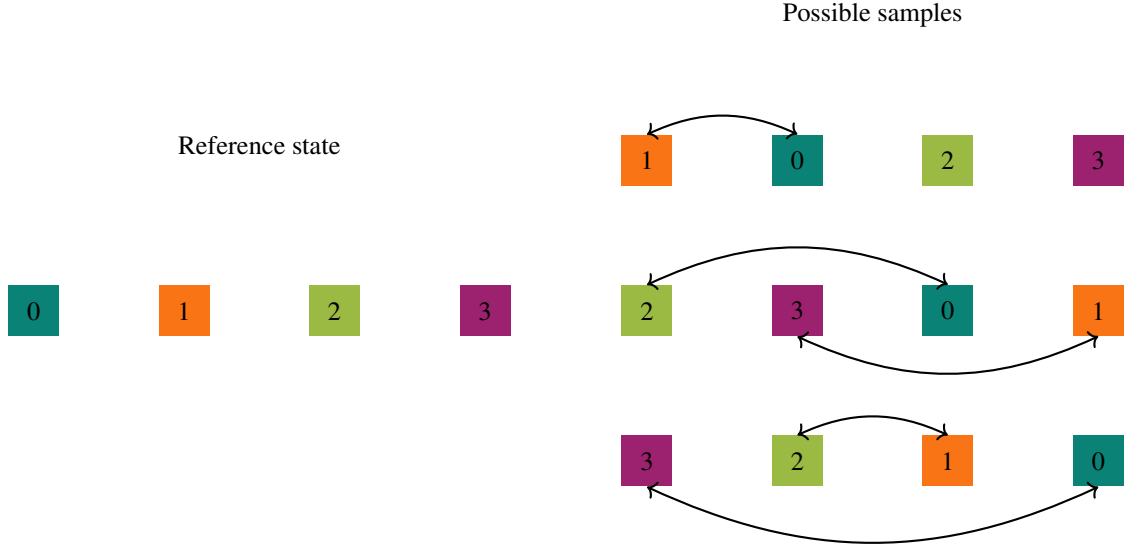
\begin{figure}[t!]
    \centering
    \begin{tikzpicture}[square/.style={regular polygon,regular polygon sides=4, minimum size=0.75cm}, align=center,node distance=2.0cm and 1.0cm, baseline=(1.base)] 
\node[square, fill=infineon_dark_green] (1) {0}; 
\node[square, fill=infineon_orange] (2) [right of=1]{1}; 
\node[square, fill=infineon_light_green] (3) [right of=2]{2}; 
\node[square, fill=infineon_purple] (4) [right of=3]{3};
\node[above=of $(2.south)!0.5!(3.south)$, yshift=0.3cm] (5) {Reference state};

\end{tikzpicture}
\hspace{1.3cm}
\begin{tikzpicture}[square/.style={regular polygon,regular polygon sides=4, minimum size=0.75cm}, align=center,node distance=2.0cm and 1.0cm, baseline=(12.base)] 
\node[square, fill=infineon_orange] (8) {1};
\node[square, fill=infineon_dark_green] (9) [right of =8] {0};
\node[square, fill=infineon_light_green] (10) [right of =9] {2};
\node[square, fill=infineon_purple] (11) [right of =10] {3};

\node[square, fill=infineon_light_green] (12) [below of =8] {2};
\node[square, fill=infineon_purple] (13) [right of =12] {3};
\node[square, fill=infineon_dark_green] (14) [right of =13] {0};
\node[square, fill=infineon_orange] (15) [right of =14] {1};

\node[square, fill=infineon_purple] (16) [below of =12] {3};
\node[square, fill=infineon_light_green] (17) [right of =16] {2};
\node[square, fill=infineon_orange] (18) [right of =17] {1};
\node[square, fill=infineon_dark_green] (19) [right of =18] {0};

\draw[<->, thick, bend left=25] (8.north) to (9.north);
\draw[<->, thick, bend left=25] (12.north) to (14.north);
\draw[<->, thick, bend right=25] (13.south) to (15.south);
\draw[<->, thick, bend left=25] (17.north) to (18.north);
\draw[<->, thick, bend right=25] (16.south) to (19.south);

\node[above=of $(9.south)!0.5!(10.south)$] (6) {Possible samples};
\end{tikzpicture}
    \caption{Example for sampling a new state from the reference on the left. Here, each color corresponds to a different node in the TSP. The exchange chain starts at index $0$ and has a length of $2$. This means the first two nodes in the route are swapped out for different ones.}
    \label{fig:sampling}
\end{figure}
Starting with the first step in the TSP, we create a Dicke state~\cite{dicke_states} with Hamming weight $1$ among all nodes except the reference node. This operation ensures a quantum state that maintains a one-hot encoding for that TSP step. While we do not allow the algorithm to sample exactly the reference state, the algorithm can freely choose any other node in the graph instead. Then, we use the last step as a buffer to store the chosen node as well as the node and time step with which it has been switched. Those two steps of the TSP are then fixed and cannot be changed in the remaining circuit. Afterwards, this process is repeated until the end of the exchange chain is reached. Since each swap fixes two nodes in the route, this algorithm only allows for exchanges chains of a length up to $\lfloor \frac{n}{2} \rfloor$ for a TSP with $N$ nodes. If the chain length exceeds this limit, nodes would have to be changed after already being fixed, which is not allowed. An example of the sampling possibilities is shown in Fig. \ref{fig:sampling}. 
For illustration purposes, we defined an example for this circuit in \href{https://algassert.com/quirk#circuit={%22cols%22:[[1,%22~94a8%22,1,1,1,1,1,1,1,1,1,1,%22X%22,%22X%22,%22X%22,%22X%22],[%22Chance4%22,1,1,1,%22Chance4%22,1,1,1,%22Chance4%22,1,1,1,%22Chance4%22],[1,%22%E2%80%A2%22,1,1,1,1,1,1,1,1,1,1,1,%22X%22],[1,1,%22%E2%80%A2%22,1,1,1,1,1,1,1,1,1,1,1,%22X%22],[1,1,1,%22%E2%80%A2%22,1,1,1,1,1,1,1,1,1,1,1,%22X%22],[%22Chance4%22,1,1,1,%22Chance4%22,1,1,1,%22Chance4%22,1,1,1,%22Chance4%22],[1,%22%E2%80%A2%22,1,1,%22X%22],[1,1,1,1,1,1,1,1,1,1,1,%22X%22,1,%22%E2%80%A2%22],[1,1,1,1,1,1,%22Swap%22,1,1,1,1,%22Swap%22,1,1,%22%E2%80%A2%22],[1,1,1,1,1,1,1,1,1,1,%22%3C%3C2%22,1,1,1,%22%E2%80%A2%22],[1,1,1,1,1,1,1,%22Swap%22,1,1,1,%22Swap%22,1,1,1,%22%E2%80%A2%22],[1,1,1,1,1,1,1,1,1,1,%22%3C%3C2%22,1,1,1,1,%22%E2%80%A2%22],[%22Chance4%22,1,1,1,%22Chance4%22,1,1,1,%22Chance4%22,1,1,1,%22Chance4%22],[1,1,1,1,%22%E2%80%A2%22,1,1,1,1,1,1,1,%22X%22],[1,1,1,1,1,1,%22%E2%80%A2%22,1,1,1,1,1,1,1,%22X%22],[1,1,1,1,1,1,1,%22%E2%80%A2%22,1,1,1,1,1,1,1,%22X%22],[%22Chance4%22,1,1,1,%22Chance4%22,1,1,1,%22Chance4%22,1,1,1,%22Chance4%22],[1,1,%22%E2%80%A2%22,1,1,1,1,1,%22X%22],[1,1,1,1,1,1,%22%E2%80%A2%22,1,1,%22X%22],[1,1,1,1,1,1,1,1,1,1,%22X%22,1,1,1,%22%E2%80%A2%22],[%22Chance4%22,1,1,1,%22Chance4%22,1,1,1,%22Chance4%22,1,1,1,%22Chance4%22],[1,1,1,1,1,1,1,1,%22%E2%80%A2%22,1,1,1,%22X%22],[1,1,1,1,1,1,1,1,1,%22%E2%80%A2%22,1,1,1,%22X%22],[1,1,1,1,1,1,1,1,1,1,%22%E2%80%A2%22,1,1,1,%22X%22],[1,1,1,1,1,1,1,1,1,1,1,%22%E2%80%A2%22,1,1,1,%22X%22],[%22Chance4%22,1,1,1,%22Chance4%22,1,1,1,%22Chance4%22,1,1,1,%22Chance4%22]],%22gates%22:[{%22id%22:%22~590m%22,%22name%22:%22%E2%88%9A1/3%22,%22matrix%22:%22{{%E2%88%9A%E2%85%93,-%E2%88%9A%E2%85%94},{%E2%88%9A%E2%85%94,%E2%88%9A%E2%85%93}}%22},{%22id%22:%22~bo2r%22,%22name%22:%22W4%22,%22circuit%22:{%22cols%22:[[%22X%22],[%22%E2%80%A2%22,%22H%22],[%22X%22,%22%E2%80%A2%22,%22H%22],[1,%22X%22,%22%E2%80%A2%22],[%22%E2%80%A2%22,1,1,%22H%22],[%22X%22,1,1,%22%E2%80%A2%22]]}},{%22id%22:%22~94a8%22,%22name%22:%22W3%22,%22circuit%22:{%22cols%22:[[%22X%22],[%22%E2%80%A2%22,%22~590m%22],[%22X%22,%22%E2%80%A2%22,%22H%22],[1,%22X%22,%22%E2%80%A2%22]]}},{%22id%22:%22~mv19%22,%22name%22:%22W2%22,%22circuit%22:{%22cols%22:[[%22X%22],[%22%E2%80%A2%22,%22H%22],[%22X%22,%22%E2%80%A2%22]]}},{%22id%22:%22~bvbm%22,%22name%22:%22W2'%22,%22circuit%22:{%22cols%22:[[%22X%22],[%22%E2%80%A2%22,1,%22H%22],[%22X%22,1,%22%E2%80%A2%22]]}},{%22id%22:%22~plrr%22,%22name%22:%22W2''%22,%22circuit%22:{%22cols%22:[[%22X%22],[%22%E2%80%A2%22,1,1,%22H%22],[%22X%22,1,1,%22%E2%80%A2%22]]}}]}}{Quirk}. 
This example performs the state preparation for a TSP with $4$ nodes and an exchange chain of length $2$. The reference solution is the path $(0,1,2,3)$.

\section{Further algorithm settings}\label{sec:alg_settings}

\subsection{Parameters in GAS}
Certain parameters in GAS only become relevant once the optimization problem is sufficiently large. One example is the choice of $\lambda$ in \Cref{alg:Grover Adaptive Search} or the termination criterion. 
In~\cite{Gilliam_2021_GroverAdaptiveSearch}, the GAS is applied to portfolio optimization for three assets, i.e., three qubits. They choose the termination criterion to stop after three consecutive iterations without improvement. Now, for larger problems, the question of good termination criteria (\textcolor{blue}{blue values} in \Cref{alg:Grover Adaptive Search highlighted}) and iteration strategies for an unknown number of marked states (\textcolor{orange}{orange values} in \Cref{alg:Grover Adaptive Search highlighted}) arises. 
\begin{algorithm}[H]
\caption{Grover Adaptive Search with steps subject to variation colored \textcolor{blue}{blue} and \textcolor{orange}{orange}.}
\label{alg:Grover Adaptive Search highlighted}

\KwData{$f: X \rightarrow \mathbb{R}$, $\lambda>1$} 
\KwResult{Solution candidate $x \in X, $ value $y=f(x)$} 
Compute $x_0$ from a heuristic and set $x=x_0, y=f(x_0)$\;
Set $r=1$\;
\While{\textcolor{blue}{Termination condition is not met}}    
{\textcolor{orange}{Randomly select $n_{\text{grover}}$ from $\{0, \dots ,\lceil\lambda^r\rceil \}$};
Apply Grover with $n_{\text{grover}}$ iterations of oracle $O_{y}$ and diffusion operator D\; 
Measure solution candidate $z$\;
\If{$f(z) < y$}
    {Set new threshold $y=f(z)$ and incumbent solution $x=z$\;
    Reset $r=1$\;}
\Else{Set $r=r+1$\;}
}

\end{algorithm}

\subsubsection{Termination criterion}
The termination criterion directly influences the gate count and the success probability of the algorithm. In this work, we investigate the following three options for termination criteria based on gate count:
\begin{itemize}
    \item Constant termination: Stop after $c$ iterations, where $c$ is a constant factor.
    \item Linear termination: Stop after $c\cdot N$ applications of Grover iterations, where $N$ is the size of the input problem.
    \item Quadratic termination: Stop after $c\cdot N^2$ applications of Grover iterations.
    
\end{itemize}
Additionally, we compare the performance of GAS with termination criteria based on the success probability. The counter $r$ in \Cref{alg:Grover Adaptive Search highlighted} counts the number of times we measure an unmarked state. Thus, allowing for larger $r$, decreases the probability of terminating even though there still is a marked state. We investigate the following options:
\begin{itemize}
    \item Constant termination: Stop when $r=c$, where $c$ is a constant factor.
    \item Linear termination: Stop when $r = c\cdot N$, where $N$ is the size of the input problem.
    \item Quadratic termination: Stop when $ r= c\cdot N^2$.
\end{itemize}

\subsubsection{Number of Grover iterations for an unknown number of marked states}\label{sec: Number of Grover iterations for an unknown number of marked states}
For applying Grover to optimization problems, we mark states in the search space dependent on a bound on the goal function. In general, for a given bound, we do not know the number of marked states. Now, if we pick the wrong number of iterations, Grover's search algorithm can result in a quantum state, where the amplitude of our desired state is low. To avoid that, we need good strategies to set the number of Grover iterations to increase our chances of finding a marked state. In this work, we investigate the following options for setting the number of iterations in GAS:
\begin{itemize}
    \item[I.] The original strategy from~\cite{Gilliam_2021_GroverAdaptiveSearch}: Randomly select the number of Grover iterations from $\{0, \dots ,\lambda^r \}$ with $\lambda > 1$ in each round $r$. \label{GAS: original strategy}
    \item[II.] Fixed interval strategy: Randomly select the number of Grover iterations from $\left\{0, \dots , \left\lceil \frac{\pi}{4 \arcsin\left(\sqrt{1/N}\right)} \right\rceil -1\right\}$. \label{GAS: fixed interval strategy}
    \item[III.] Incremental number of iterations: Fix the number of iterations to $\left\lfloor  \frac{\pi}{4 \arcsin\left(\sqrt{2^{-r+1}}\right)} \right\rfloor$ for each round $r$. \label{GAS: incremental strategy}
\end{itemize}
The original strategy (\hyperref[GAS: original strategy]{I.}) is outlined in \Cref{alg:Grover Adaptive Search}. Initially, the number of iterations is selected uniformly at random from a small interval. This interval systematically expands as long as a marked state is not measured. The process terminates upon the measurement of a marked state or when a predefined termination criterion (most commonly a maximum number of Grover iterations) is met. Essentially, the strategy begins with a limited number of iterations and progressively increases this allowance if unmarked states are repeatedly measured. This iterative adjustment allows Strategy \hyperref[GAS: original strategy]{I} to achieve a success probability that can be arbitrarily close to 1, depending on the chosen termination condition. 
 
The fixed interval strategy (\hyperref[GAS: fixed interval strategy]{II.}) is an attempt to maximize the success probability. We choose the set from which we draw the number $n_{\text{grover}}$ of Grover iterations to be $\{0, \dots , \lceil \frac{\pi}{4 \Theta} \rceil -1\}$ with $\Theta = \arcsin\left(\sqrt{\frac{1}{N}}\right)$. As stated in Lemma 2 in~\cite{Boyer_1998_tightBoundsQuantumSearch}, when applying $n_{\text{grover}}$ Grover iterations to the uniform superposition, where $n_{\text{grover}}$ is uniformly randomly drawn from the set $\{0,\dots , m-1 \}$ with $m\geq 1/\sin{2\Theta}$ and $\Theta = \arcsin{\sqrt{M/N}}$, the probability of obtaining a marked state is at least $1/4$. For small values of $\Theta$ we have $\sin{2\Theta} \approx 2\Theta$. Thus, our upper bound of the set in the fixed interval strategy $\frac{\pi}{4\arcsin\left(\sqrt{1/N}\right)}-1$ also ensures the probability of measuring a marked state to be at least $1/4$ per round, since it fulfills
\begin{align}
    \frac{1}{\sin{2\Theta}} \approx \frac{1}{2\Theta}= \frac{1}{2\arcsin{\sqrt{M/N}}}\leq \frac{1}{2\arcsin{\sqrt{1/N}}} \leq \frac{\pi}{4\arcsin{\sqrt{1/N}}}.
\end{align}
 
The incremental strategy (\hyperref[GAS: incremental strategy]{III.}) is inspired by the lecture notes by Scott Aaronson~\cite{Aaronson_2018_LectureNotes}. In Strategy \hyperref[GAS: incremental strategy]{III.}, we estimate the number of marked states in each round. If no marked state is measured, this guess is updated. Specifically, the protocol begins with an initial guess for the number of marked items $M=N$ and halves this guess in subsequent steps. This halving mechanism means that fewer Grover iterations are applied initially, with the number of iterations increasing exponentially in later stages. Unlike protocols that randomly draw the number of queries, the incremental strategy (\hyperref[GAS: incremental strategy]{III.}) deterministically updates its guess in each round until either a marked state is found or a predefined termination condition is met.

\section{Results and Discussion}

We analyze $15$ TSP instances with $5$ randomly generated problems with $8$, $10$ and $12$ nodes respectively, the number of decision variables hence lies between $64$ and $144$. For each instance, we compute the initial objective value via a simple greedy procedure: Starting at an arbitrary node, we add the shortest available edge leading to an unvisited node to the tour until all nodes have been visited. Then, we use the classical optimization library OR-Tools, specifically, the CP-SAT solver~\cite{cpsatlp} to compile the set
\begin{equation}\label{eq:good states}
    F_T = \{x \in \{0,1\}^{n \times  n}: x \text{ is a feasible tour, } d(x) < T\},
\end{equation}
where $n$ is the number of nodes, $d(x)$ is the distance of a tour, and $T$ is the initial objective threshold.
With this information, we can classically simulate our Grover-based protocols without the need to run quantum circuits. This is done by (randomly) drawing the number of Grover iterations $n_{\text{grover}}$ in each round and computing the measurement for each of the previously computed ``good" states according to the formula
\begin{equation}
    p_j(x) = p_0(x) \frac{\sin^2((2j+1)\arcsin(\sqrt{P}))}{P}.
\end{equation}
Here $M=|F_T|$ is the number of good states, $P=\frac{M}{n!}$ is the sum of all the measurement probabilities belonging to good states, and $p_0(x) = \frac{1}{n!}$ is the initial measurement probability for state $x$. This method of evaluating a Grover-based algorithm works as long as the values of $p_0(x)$ and $P$ can be efficiently obtained which is the case as long as the computation of eq. \ref{eq:good states} takes a reasonable amount of time.
We found that this was still the case for the instance sizes under consideration which at over $100$ variables go much beyond the scope of any benchmarking attempts for other quantum algorithms such as QAOA.
As a termination condition, we limited the number of rounds within each improvement step to $R=5$, $R=\log_{\lambda}(n^2)$ or $R=\log_{\lambda}(n^4)$, respectively, where $\lambda = 5/4$ is the base of the exponentially growing upper bound for the number of Grover iterations.
Recall that the number of Grover iterations is sampled uniformly at random from the interval $[0,\lceil \lambda^k \rceil]$ in round $k$. Hence the expected (and maximum) number of iterations is $\mathcal{O}(\lambda^R)$ if we allow for $R$ rounds. As our highest possible choice of $R$ was $\log_{\lambda}(n^4)$, our algorithm overall has a polynomial run time.

\begin{figure}[H]
    \centering
    \includegraphics[width=0.48\linewidth]{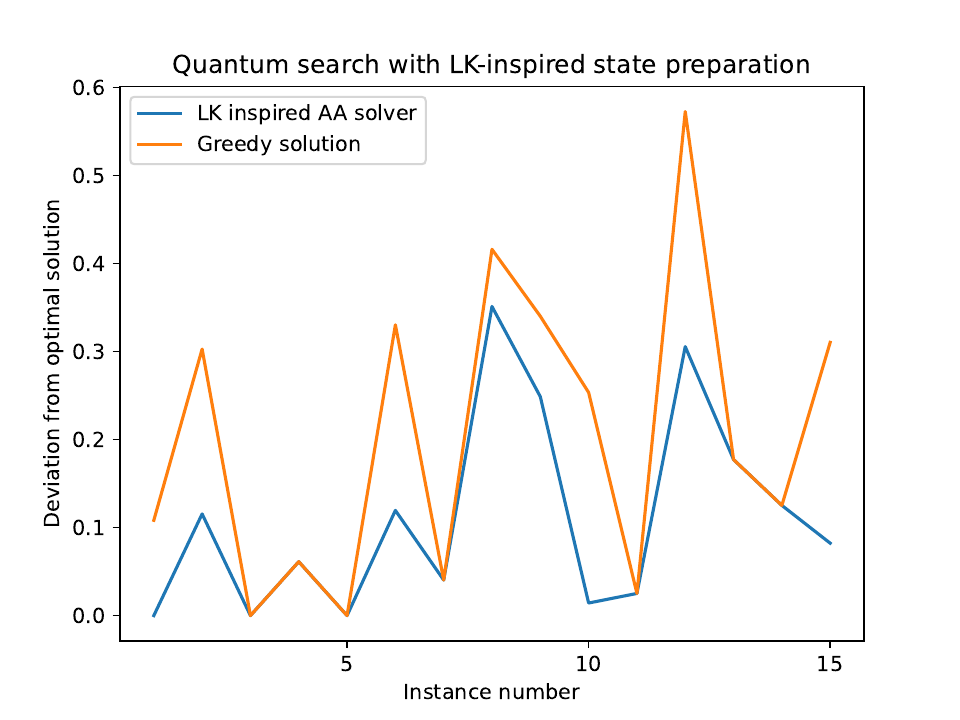}
    \includegraphics[width=0.48\linewidth]{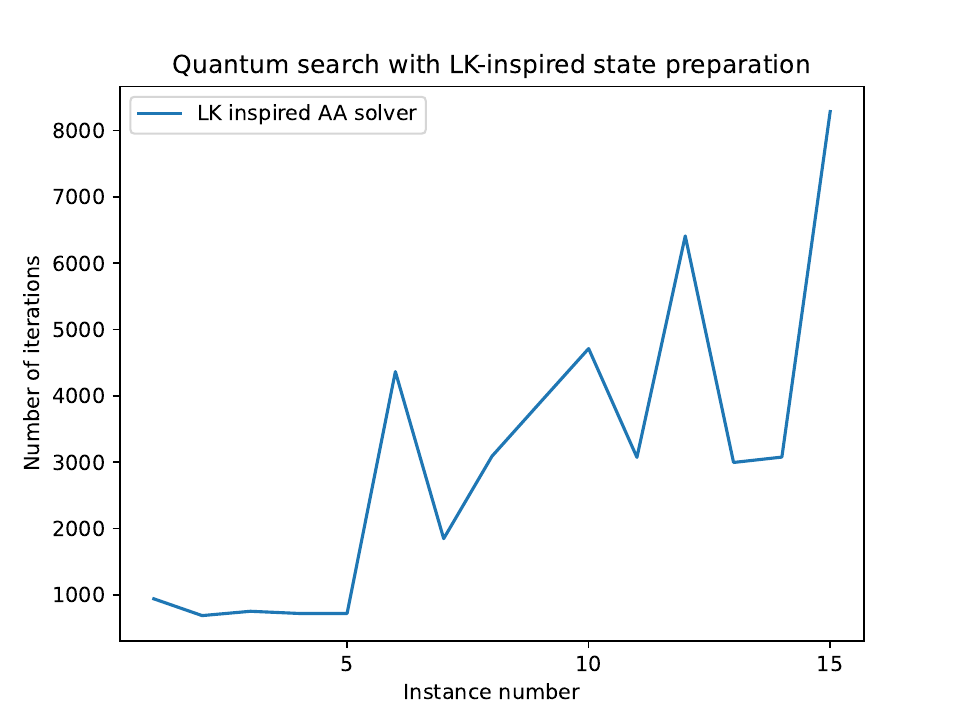}

    \caption{Approximation ratio and iteration count of the local search strategy inspired by the Lin-Kernighan algorithm.}
    \label{fig:approximation_ratio}
\end{figure}
To measure the performance of our proposed methods, we introduce two metrics. The first performance metric is the deviation of the detected solution from the optimal one, averaged over a number of trial runs.
Concretely, if $x$ is the detected solution and $x^*$ is the optimal solution, our figure of merit is given by
\begin{equation}
    \frac{d(x)-d(x^*)}{d(x^*)}.
\end{equation}
The second performance measure in this work is the number of Grover iterations needed in the simulation runs to reach our results. 

The presented results go in two directions. First we briefly benchmark a quantum search algorithm with the ``local search" state preparation routine outlined in section~\ref{subsec:state_prep}.
For the instances under consideration, the ability to move away from the greedy solution with local steps proved to be limited. We suspect that this is the case because the changes required to improve the intermediate solutions would require changing almost all of the (few) assignments in the relatively small examples considered. We expect to see a more favorable behavior for larger instances, where it is possible to improve a given solution locally.
We ran the algorithm with an iteration limit of $5n^3$ where $n=8,10,12$ is the number of vertices in the graph.

Next, we examine a quantum search algorithm on the same TSP instances starting from a superposition of all feasible states with a focus on comparing the different variants of Grover adaptive search presented in section~\ref{sec:alg_settings}. 
With a termination criterion, which limits the number of rounds in an improvement step, the three strategies show some differences. Although the fixed interval and incremental versions reach a higher approximation ratio for the same number of rounds before the search is terminated, their total number of iterations is also much higher.
For the fixed interval strategy, this is because the upper limit of the interval from which we draw the number of Grover iterations does not grow throughout the protocol but starts off at a high value. For the incremental strategy, the number of Grover iterations in round $k$ scales with $\sqrt{2^k}$, compared to $\sqrt{L^k}$ for the standard GAS strategy. With our choice of $L=\frac{5}{4}$, this leads to a much higher number of Grover iterations when the round limit is chosen to be high.

\begin{figure}[h!]
    \centering
    \includegraphics[width=0.49\linewidth]{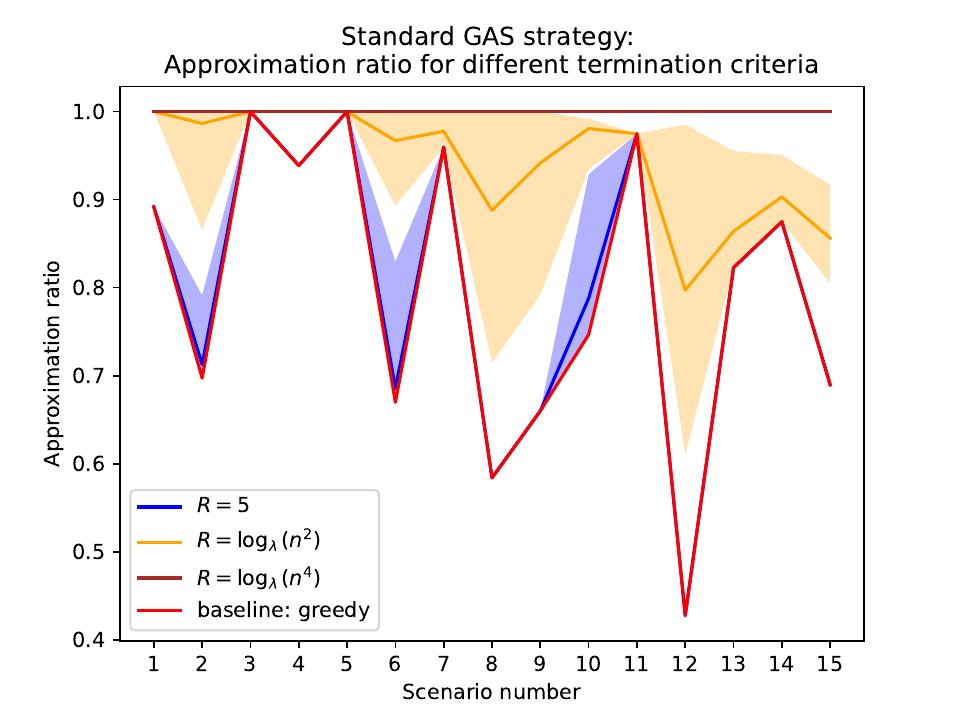}
    \includegraphics[width=0.49\linewidth]{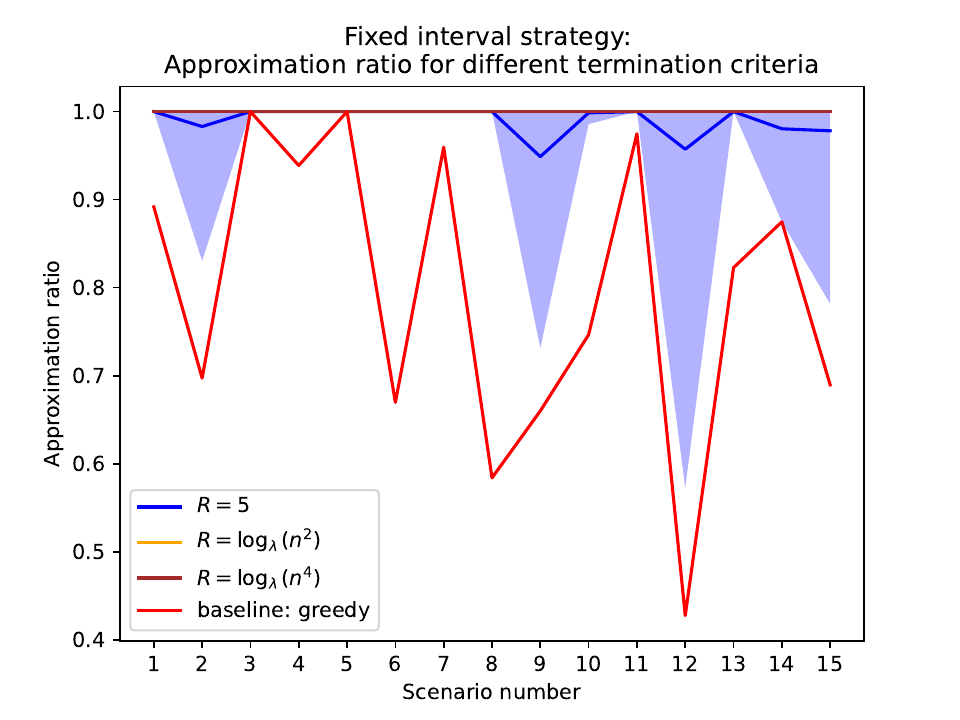}
    \includegraphics[width=0.49
    \linewidth]{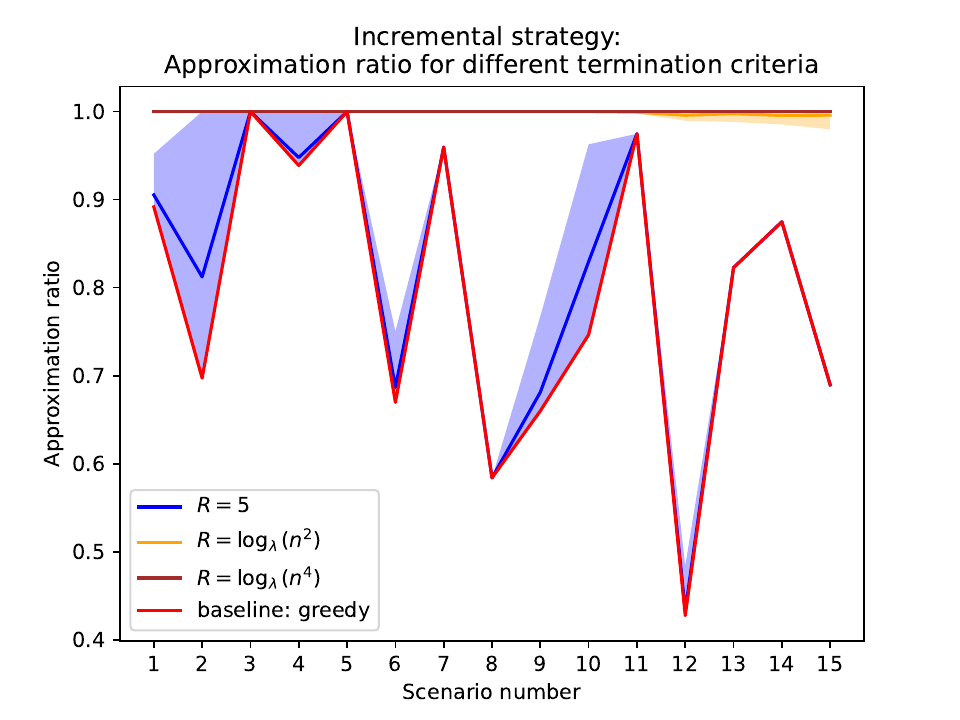}
    \caption{Comparison of the three variations of GAS outlined in \Cref{alg:Grover Adaptive Search highlighted}. We show the approximation ratio for three different choices for the number of Grover iterations, cf. \Cref{sec: Number of Grover iterations for an unknown number of marked states}. Additionally, we vary the termination criteria, where the upper bound for the number of Grover iterations is set by $\lambda^R$ and $R$ is the number of rounds within each improvement step.}
    \label{fig:approximation_ratio}
\end{figure}

\begin{figure}[h!]
    \centering
    \includegraphics[width=0.48\linewidth]{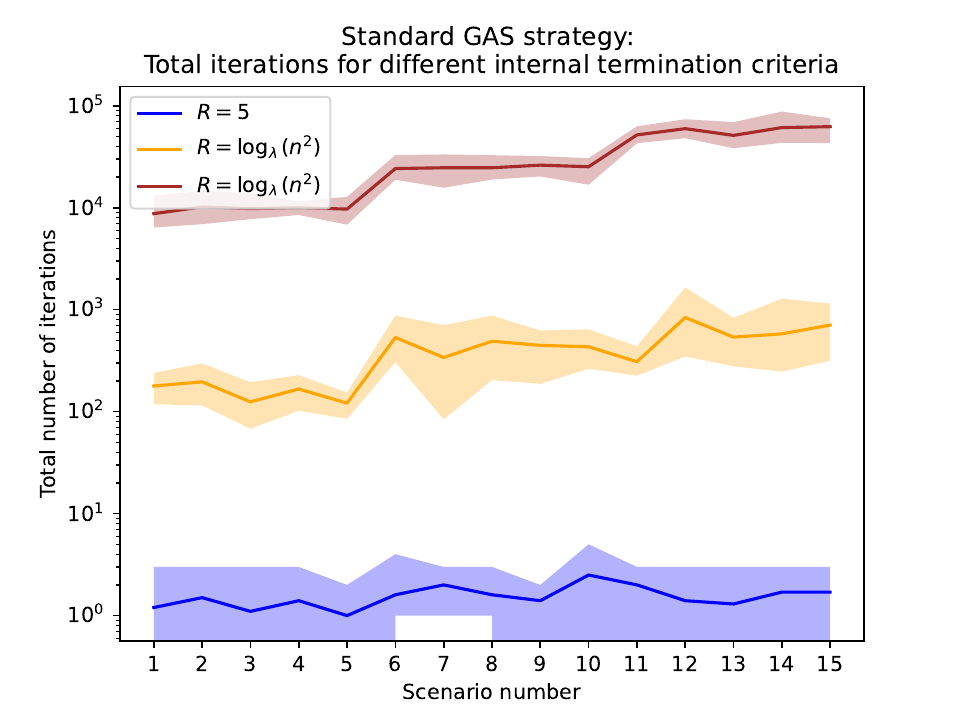}
    \includegraphics[width=0.48\linewidth]{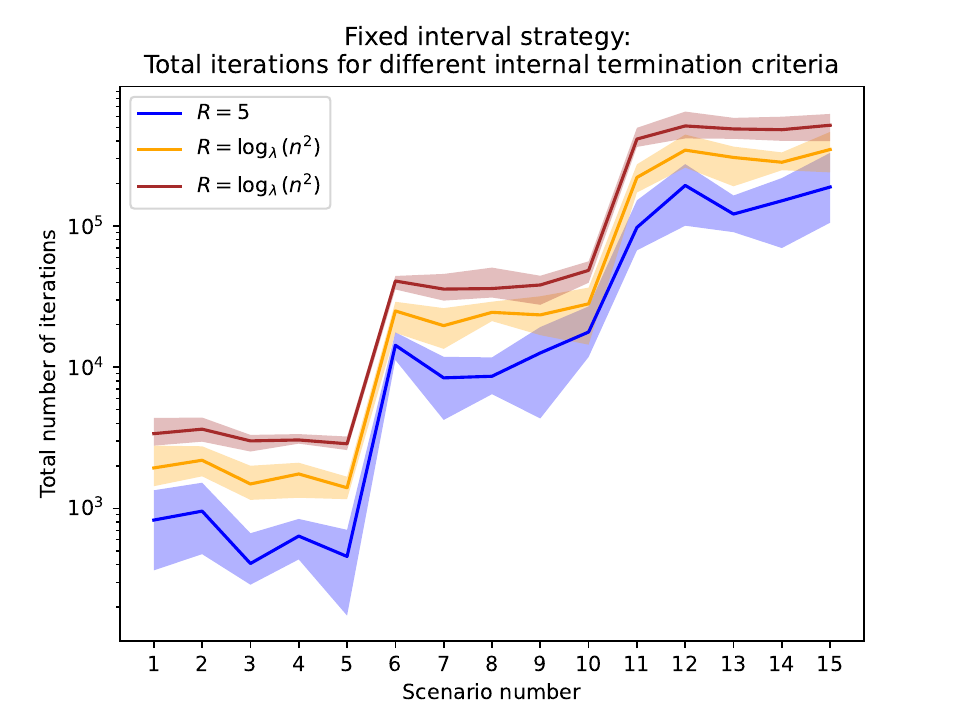}
    \includegraphics[width=0.48\linewidth]{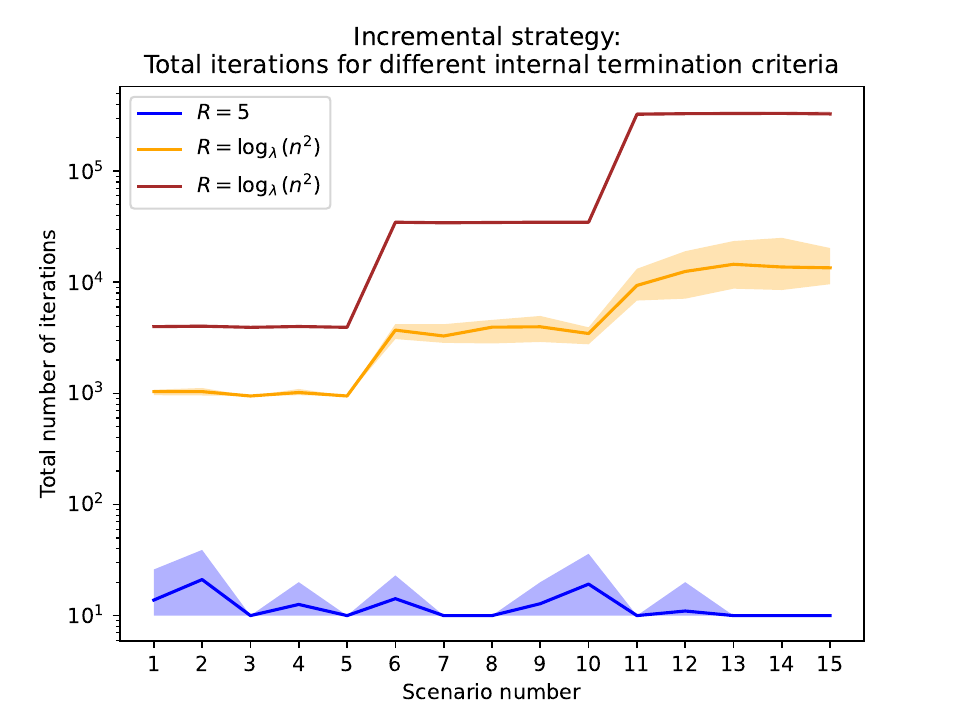}

    \caption{Comparison of the three variations of GAS outlined in~\Cref{alg:Grover Adaptive Search highlighted}. We show the total number of iterations for three different choices for the number of Grover iterations, cf.~\Cref{sec: Number of Grover iterations for an unknown number of marked states}. The termination criteria depend on the upper bound for the number of Grover iterations $\lambda^R$, where $R$ is the number of rounds within each improvement step.}
    \label{fig:iterations}
\end{figure}

\newpage
\section{Conclusion and Outlook}
Regarding the analysis of different algorithmic settings relating to choosing the number of Grover iterations, we have come to a number of conclusions.
First, it should be noted that the fixed interval strategy is not to be recommended as a general purpose termination criterion. It can be useful when the number of marked states is expected to be very low as we then skip the regime of ``few" Grover iterations and already have a reasonable probability for measuring a good state starting from the first round. As the number of marked states, or even its rough size, is generally unknown, we advise to employ the fixed interval strategy only in select cases where one has some prior knowledge.
It is not trivial to pick a winner from the standard GAS and incremental strategies. It should be noted that the number of iterations goes up much faster for the incremental strategy as it is governed by exponential number with base $2$ by construction. This growth can be controlled somewhat better by employing the standard GAS strategy and choosing the algorithm parameter $L$ to be smaller than $2$. Also, as the number of iterations is chosen deterministically, the incremental strategy is only guaranteed to eventually result in success if the amplitude of the good states is of the form $\frac{k}{N}$ where $k>0$ is an integer and $N$ is the number of solutions in the initial state. This was the case for our TSP example but might not always be true as evidenced by works like~\cite{wilkening_2023_QTGKnapsack} where the amplitudes are determined by elaborate biasing procedures.
To determine whether fixed-point quantum search~\cite{fixed_point_quantum_search} can offer an advantage in practice by replacing the randomness overhead with an algorithmic overhead is another interesting question meriting its own numeric study.

The state preparation routine preceding the search plays a key role in making amplitude amplification a useful tool in combinatorial optimization. We built on the routine presented in~\cite{state_preparation} and showed how it can be modified to a neighborhood search state. It remains to be seen whether a quantum search based algorithm employing this neighborhood search scheme can be competitive with classical heuristics. This paper cannot yet precisely answer this question as our benchmarking scheme relied on computing the sets of improving solutions (eq. ~\ref{eq:good states}). We expect that the instances for which the proposed neighborhood search is effective are larger than those we could analyze with our benchmarking scheme.
Another topic which we do not address in this work is the quantum circuits required to implement oracles. A general implementation for QUBO problems is given in~\cite{grover_adaptive_search} but more efficient problem-specific variants might exist.

\bibliographystyle{unsrt}
\bibliography{references}

@article{Gilliam_2021_GroverAdaptiveSearch,
   title={Grover Adaptive Search for Constrained Polynomial Binary Optimization},
   volume={5},
   ISSN={2521-327X},
   url={http://dx.doi.org/10.22331/q-2021-04-08-428},
   DOI={10.22331/q-2021-04-08-428},
   journal={Quantum},
   publisher={Verein zur Forderung des Open Access Publizierens in den Quantenwissenschaften},
   author={Gilliam, Austin and Woerner, Stefan and Gonciulea, Constantin},
   year={2021},
   month=apr, pages={428} }

@article{wilkening_2023_QTGKnapsack,
   title={A quantum algorithm for solving 0-1 Knapsack problems},
   volume={11},
   ISSN={2056-6387},
   url={http://dx.doi.org/10.1038/s41534-025-01097-8},
   DOI={10.1038/s41534-025-01097-8},
   number={1},
   journal={npj Quantum Information},
   publisher={Springer Science and Business Media LLC},
   author={Wilkening, Soeren and Lefterovici, Andreea-Iulia and Binkowski, Lennart and Perk, Michael and Fekete, Sandor P. and Osborne, Tobias J.},
   year={2025},
   month={aug} }

@misc{durr1999quantumalgorithmfindingminimum,
      title={A Quantum Algorithm for Finding the Minimum}, 
      author={Christoph Durr and Peter Hoyer},
      year={1999},
      eprint={quant-ph/9607014},
      archivePrefix={arXiv},
      primaryClass={quant-ph},
      url={https://arxiv.org/abs/quant-ph/9607014}, 
}

@misc{Aaronson_2018_LectureNotes,
  author        = {Scott Aaronson},
  title         = {Introduction to Quantum Information Science Lecture Notes},
  month         = {Fall},
  year          = {2018},
  url           = {https://www.scottaaronson.com/qclec.pdf}
}

@article{grover_adaptive_search,
  title={Grover adaptive search for constrained polynomial binary optimization},
  author={Gilliam, Austin and Woerner, Stefan and Gonciulea, Constantin},
  journal={Quantum},
  volume={5},
  pages={428},
  year={2021},
  publisher={Verein zur F{\"o}rderung des Open Access Publizierens in den Quantenwissenschaften}
}

@article{fixed_point_quantum_search,
  title={Fixed-point quantum search with an optimal number of queries},
  author={Yoder, Theodore J and Low, Guang Hao and Chuang, Isaac L},
  journal={Physical review letters},
  volume={113},
  number={21},
  pages={210501},
  year={2014},
  publisher={APS}
}

@inproceedings{dicke_states,
  title={Deterministic preparation of Dicke states},
  author={B{\"a}rtschi, Andreas and Eidenbenz, Stephan},
  booktitle={International Symposium on Fundamentals of Computation Theory},
  pages={126--139},
  year={2019},
  organization={Springer}
}

@inproceedings{state_preparation,
  title={Grover mixers for QAOA: Shifting complexity from mixer design to state preparation},
  author={B{\"a}rtschi, Andreas and Eidenbenz, Stephan},
  booktitle={2020 IEEE International Conference on Quantum Computing and Engineering (QCE)},
  pages={72--82},
  year={2020},
  organization={IEEE}
}

@article{grover_algorithm,
  title={Quantum computers can search arbitrarily large databases by a single query},
  author={Grover, Lov K},
  journal={Physical review letters},
  volume={79},
  number={23},
  pages={4709},
  year={1997},
  publisher={APS}
}

@article{lin_kernighan,
  title={An effective heuristic algorithm for the traveling-salesman problem},
  author={Lin, Shen and Kernighan, Brian W},
  journal={Operations research},
  volume={21},
  number={2},
  pages={498--516},
  year={1973},
  publisher={Informs}
}

@article{Boyer_1998_tightBoundsQuantumSearch,
   title={Tight Bounds on Quantum Searching},
   volume={46},
   ISSN={1521-3978},
   url={http://dx.doi.org/10.1002/(SICI)1521-3978(199806)46:4/5<493::AID-PROP493>3.0.CO;2-P},
   DOI={10.1002/(sici)1521-3978(199806)46:4/5<493::aid-prop493>3.0.co;2-p},
   number={4–5},
   journal={Fortschritte der Physik},
   publisher={Wiley},
   author={Boyer, Michel and Brassard, Gilles and Høyer, Peter and Tapp, Alain},
   year={1998},
   month=jun, pages={493–505} }

@article{Lu_2025_GASforVehicleRouting,
author = {Lu, Liu and Qian, Ling and Xy, Wu and Fan, Chen-Rui and Zhang, Lu-Fan and Cai, Dun-Bo and Lu, Hui-Jun and Wang, Tie-Jun and Wang, Chuan},
year = {2025},
month = {01},
pages = {1-11},
title = {Solving Vehicle Routing Problem Using Grover Adaptive Search Algorithm},
volume = {PP},
journal = {IEEE Transactions on Intelligent Transportation Systems},
doi = {10.1109/TITS.2025.3562860}
}

@article{Cade_2023,
   title={Quantifying Grover speed-ups beyond asymptotic analysis},
   volume={7},
   ISSN={2521-327X},
   url={http://dx.doi.org/10.22331/q-2023-10-10-1133},
   DOI={10.22331/q-2023-10-10-1133},
   journal={Quantum},
   publisher={Verein zur Forderung des Open Access Publizierens in den Quantenwissenschaften},
   author={Cade, Chris and Folkertsma, Marten and Niesen, Ido and Weggemans, Jordi},
   year={2023},
   month=oct, pages={1133} }

@misc{farhi2014quantumapproximateoptimizationalgorithm,
      title={A Quantum Approximate Optimization Algorithm}, 
      author={Edward Farhi and Jeffrey Goldstone and Sam Gutmann},
      year={2014},
      eprint={1411.4028},
      archivePrefix={arXiv},
      primaryClass={quant-ph},
      url={https://arxiv.org/abs/1411.4028}, 
}

@book{historyTSP,
  title={Graph Theory, 1736-1936},
  author={Biggs, Norman and Lloyd, E Keith and Wilson, Robin J},
  year={1986},
  publisher={Oxford University Press}
}

@misc{history2TSP,
      title={Der Handlungsreisende wie er sein soll und was er zu thun hat, um Auftraege zu erhalten und eines gluecklichen Erfolgs in seinen Geschaeften gewiss zu sein}, 
      author={Voigt},
      year={1832},
      url={https://zs.thulb.uni-jena.de/receive/jportal_jparticle_00248075}, 
}

@software{cpsatlp,
  title = {CP-SAT},
  version = { v9.12 },
  author = {Laurent Perron and Frédéric Didier},
  organization = {Google},
  url = {https://developers.google.com/optimization/cp/cp_solver/},
  date = { 2025-02-17 }
}

\end{document}